\documentclass[10pt]{iopart}
\usepackage{iopams}  
\usepackage{graphicx}
\usepackage{rotating}
\usepackage{xspace}

\def\thercsid{\relax}
\def\rcsid#1{\def\next##1#1{\def\thercsid{##1}}\next}
\rcsid$Id: s2-inspiral.tex,v 1.8 2005/04/04 18:22:55 jolien Exp $

\renewcommand{\today}{\number\day\space\ifcase\month\or
  January\or February\or March\or April\or May\or June\or
  July\or August\or September\or October\or November\or December\fi
  \space\number\year}

\newcommand{\datafind}{{\textsc{datafind}}\xspace}
\newcommand{\findchirp}{{\textsc{findchirp}}\xspace}
\newcommand{\inca}{{\textsc{inca}}\xspace}
\newcommand{\inspiral}{{\textsc{inspiral}}\xspace}
\newcommand{\tmpltbank}{{\textsc{tmpltbank}}\xspace}
\newcommand{\trigtotmplt}{{\textsc{trigbank}}\xspace}

\begin{document}

\title[The \inspiral program]%
{Using the \pmb{\sc inspiral} program to search for gravitational waves from
low-mass binary inspiral}

\author[Duncan A. Brown et al.]{Duncan A. Brown (for the LIGO Scientific Collaboration)}
\address{University of Wisconsin--Milwaukee, Milwaukee, WI 53211, USA}
\address{California Institute of Technology, Pasadena, CA 91125, USA}
\begin{abstract}
The \inspiral program is the LIGO Scientific Collaboration's computational
engine for the search for gravitational waves from binary neutron stars and
sub-solar mass black holes.  We describe how this program, which makes use of
the \findchirp algorithm (discussed in a companion paper), is integrated into a
sophisticated data analysis pipeline that was used in the search for low-mass
binary inspirals in data taken during the second LIGO science run.
\end{abstract}


\submitto{\CQG}


\section{Introduction}

The search for gravitational waves that arise from coalescing compact binary
systems---binary neutron stars and black holes---is one of the main efforts 
in LIGO data analysis conducted by the LIGO Scientific Collaboration (LSC).
For the first science run of LIGO, we focused attention on the search for
low-mass binary systems (binary neutron stars and sub-solar mass binary black
hole systems) since these systems have well-understood
waveforms~\cite{s1inspiral}.  For the second science run of LIGO, S2, we have
extended our analysis to include searches for binary black hole systems with
higher masses.  This work will be described elsewhere~\cite{s2bbh}.  Here we
describe the refinements to the original S1 search for low-mass binary
inspiral waveforms: binary neutron star systems~\cite{s2bns} and sub-solar-mass
binary black hole systems which may be components of the Milky Way Halo (and
thus form part of a hypothetical population of Macroscopic Astrophysical
Compact Halo Objects known as MACHOs)~\cite{s2macho}.  In this paper we focus
on how the central computational search engine---the \inspiral program---is
integrated into a sophisticated new data analysis pipeline that was used in
the S2 low-mass binary inspiral search.

The \inspiral program makes use of the LSC inspiral
search algorithm \findchirp, which is described in detail in a companion
paper~\cite{findchirp}.  This companion paper describes the algorithm that we
use to generate inspiral triggers given an inspiral template and a \emph{single
data segment}.  There is more to searching for gravitational waves from binary
inspiral than trigger generation, however.  To perform a search for a given
class of sources in a large quantity of interferometer data we construct a
\emph{detection pipeline}.

The detection pipeline is divided into blocks that perform specific tasks.
These tasks are implemented as individual programs.  The pipeline itself is
then implemented as a logical graph, called a \emph{directed acyclic graph} or
DAG, describing the workflow (the order that the programs must be called to
perform the pipeline activities) which can
then be submitted to a batch processing system on a computer cluster.
We use the Condor high throughput computing system~\cite{condor} to manage the
DAG and to control the execution of the programs.

The structure of this paper will follow these various tasks (programs).
In Sec.~\ref{s:overview} we provide an overview of these tasks and then
describe each task in turn.  In Sec.~\ref{s:pipeline} we provide a simple
illustration of how these elements are combined into a workflow pipeline.

\section{Pipeline overview}
\label{s:overview}

In LIGO's second science run S2 we performed a \emph{triggered} search for
low-mass binary inspirals: since we require that a trigger occur simultaneously
and consistently in at least two detectors located at different sites in order
for it to be considered as a detection candidate, we save computational effort
by analyzing data from the Livingston interferometer first and then performing
follow-up analyses of Hanford data only for the specific triggers found.  We
describe the tasks and their order of execution in this triggered search as
our detection pipeline.

An overview of the tasks required in the detection pipeline are: (i) data
selection with the \datafind program, as described in Sec.~\ref{ss:datafind},
(ii) template bank generation with the \tmpltbank program, as described
in Sec.~\ref{ss:tmpltbank}, (iii) binary inspiral trigger generation (the
actual filtering of the data) with the \inspiral program, as described
in Sec.~\ref{ss:inspiral}, (iv) creation of a follow-up (in other detectors)
template bank based on these triggers generated with the \trigtotmplt program,
as described in Sec.~\ref{ss:trigtotmplt}, and finally (v) coincidence analysis
of the triggers with the \emph{in}spiral \emph{c}oincidence \emph{a}nalysis
program \inca, as described in Sec.~\ref{ss:inca}.

\begin{figure}[t]
\begin{center}
\includegraphics[width=0.6\linewidth]{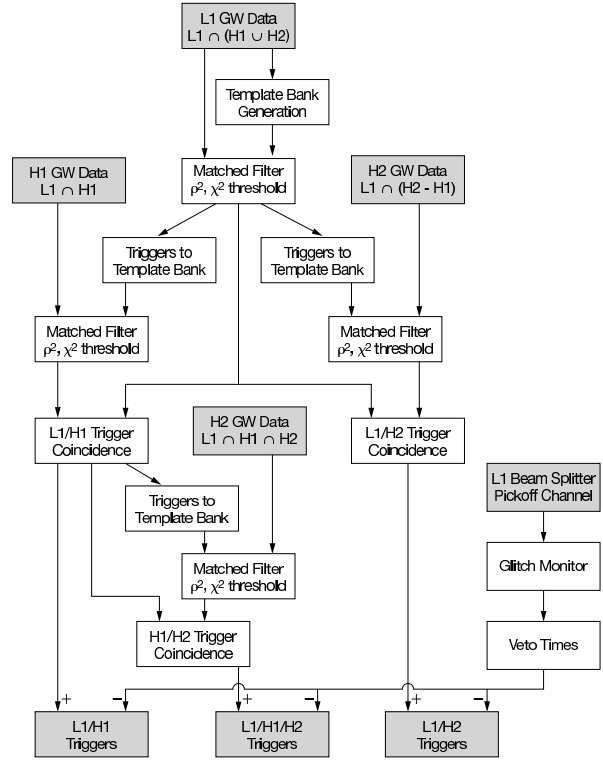}
\end{center}
\caption[Structure of the S2 Triggered Search Pipeline]{%
\label{f:pipeline}
The triggered search pipeline. $\mathrm{L1} \cap (\mathrm{H1} \cup \mathrm{H2})$ indicates times when
the L1 interferometer was operating in coincidence with one or both of the
Hanford interferometers. $\mathrm{L1} \cap \mathrm{H1}$ indicates times when
the L1 interferometer was operating in coincidence with the H1 interferometer.
$\mathrm{L1} \cap (\mathrm{H2} - \mathrm{H1})$ indicates times when the L1
interferometer was operating in coincidence with only the H2 interferometer.
The outputs of the search pipeline are triggers that belong to one of the
two double coincident data sets or to the triple coincident data set.}
\end{figure}

Figure \ref{f:pipeline} shows our workflow in terms of these basic tasks.
Epochs of simultaneous Livingston-Hanford operation are processed differently
depending on which interferometer combination is operating.  Thus, there are
several different sets of data: $\mbox{L1}\cap(\mbox{H1}\cup\mbox{H2})$
is when the Livingston interferometer L1 is operating simultaneously with
either the 4~km Hanford interferometer H1 or the 2~km Hanford interferometer H2
(or both)---this is all the data analyzed by the S2 low-mass inspiral
analysis---while $\mbox{L1}\cap\mbox{H1}$ is when L1 and H1 are both operating
operating,
$\mbox{L1}\cap(\mbox{H2}-\mbox{H1})$ is when L1 and H2 but not H1 are
operating, and $\mbox{L1}\cap\mbox{H1}\cap\mbox{H2}$ is when all three
interferometers are operating.  A full L1 template bank is generated for 
the $\mbox{L1}\cap(\mbox{H1}\cup\mbox{H2})$ data and the L1 data is filtered
with \inspiral.  Triggers resulting from this are then used to produce
\emph{triggered banks} for followup filtering of H1 and/or H2 data.  However,
if both H1 and H2 are operating then filtering of H2 is suspended until
coincident L1/H1 triggers are identified by \inca.  After a final coincidence
check, final triggers are checked to see if they could be \emph{vetoed} by
association with detector misbehaviour, which, for the S2 inspiral analysis,
was diagnosed by examining an auxiliary interferometer channel of the L1
analysis for glitches~\cite{s2bns}.

This pipeline is equivalent to simply
filtering the data from the three interferometers and looking for coincident
triggers.  The triggered search is more computationally efficient because H1
and H2 data is filtered only if it could possibly produce a coincident trigger,
i.e., only for times when there are L1 triggers produced.

\subsection{The \datafind program}
\label{ss:datafind}

\begin{figure}[t]
\begin{center}
\includegraphics[width=0.6\linewidth]{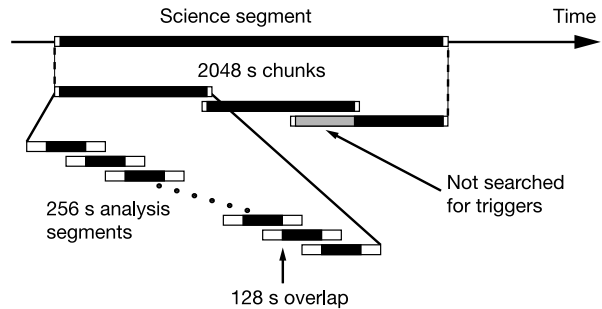}
\end{center}
\caption[Algorithm Used to Divide Science Segments into Data Analysis
Segments]{%
\label{f:s2_segments}
The algorithm used to divide science segments into data analysis segments.
Science segments are divided into $2048$~s chunks overlapped by $128$~s.
(Science segments shorter than $2048$~s are ignored.) An additional chunk with
a larger overlap is added to cover any remaining data at the end of a science
segment.  Each chunk is divided into $15$ analysis segments of length $256$~s
for filtering. The first and last $64$~s of each analysis segment is ignored,
so the segments overlap by $128$~s.  Areas shaded black are filtered for
triggers by the search pipeline. The gray area in the last chunk of the
science segment is not searched for triggers as this time is covered by the
preceding chunk, however this data is used in the PSD estimate for the final
chunk.}
\end{figure}

The \datafind program \cite{ldr} uses tables of \emph{science segments}---lists of time
when each interferometer was operating stably and producing ``science mode''
data---to identify \emph{data chunks} to be used in the analysis.  The chunks
are composed of 15 overlapping \emph{analysis segments} of 256~s which are
used for power spectral density (PSD) estimation and for matched filtering with
the \findchirp algorithm.  For the S2 inspiral analysis chunks of 2048~s were
used and are chosen so that they overlap by 128~s to allow continuous
filtering across the boundary between chunks (though the last chunk in a
science segment overlaps by a larger amount so that it can remain 2048~s long).
Figure~\ref{f:s2_segments} shows how analysis chunks are constructed from
science segments in S2.

\subsection{The \tmpltbank program}
\label{ss:tmpltbank}

The \tmpltbank program \cite{lalapps} is responsible for producing a bank of waveform
templates to be used as matched filters by the \inspiral program.  In the
triggered search pipeline the L1 data is filtered first and since the
coincidence stage requires triggers to be observed in the same template (see
below), it is sufficient to generate a template bank suitable for the L1 data.
To construct a template bank \tmpltbank needs an estimate of the average strain
noise PSD of the instrument.  This is computed for
each data analysis chunk (as provided by \datafind) using the same method as
the \inspiral code.  The specific templates that produce inspiral triggers in
the \inspiral program will then be used as the bank for subsequent analyses of
H1 and/or H2 data.

The S2 inspiral and MACHO searches looked for signals from binary systems where the
component masses were in the range $1\,M_\odot\le m_1,m_2\le 3M_\odot$ (the
binary neutron stars) and the range $0.2\,M_\odot\le m_1,m_2\le 1M_\odot$ (the
binary black hole MACHOs).  Since each mass pair $m_1,m_2$ in the space
produces a different waveform, we construct a \emph{template bank}---a discrete
subset of the continuous family of waveforms that belong to the parameter
space.  The placement of the templates in the bank is determined by the
\emph{mismatch} of the bank which is the maximum fractional loss of
signal-to-noise ratio that can occur by filtering a true signal with component
masses $m_1,m_2$ with the ``nearest'' template waveform for a system with
component masses $m_1',m_2'$.  The construction of an appropriate template
bank is discussed in Refs.~\cite{Owen:1995tm,Owen:1998dk}.

A typical choice for the maximum bank mismatch
is 3\%, which would correspond to a 10\% loss of event rate for a population of
sources with uniform spatial distribution.  This 3\% bank mismatch was adopted
in the S2 binary neutron star search.  Binary black hole MACHOs, if they exist,
would exist in the Milky Way halo and would be easily detectable; it is not
so important for the template bank to have such a small mismatch.  By raising
the mismatch, fewer templates are needed in the bank.  This is an important
computational-cost saving strategy for the binary black hole MACHO search
as the template bank is very large for these very-low mass sources.  For the
S2 binary black hole MACHO search we adopted a mismatch of 5\%.

\subsection{The \inspiral program}
\label{ss:inspiral}

The \inspiral program \cite{lalapps} is the driver for the \findchirp algorithm, which is
discussed in detail in \cite{findchirp}.  This program acquires a chunk of
data identified by \datafind and the bank of templates produced by \tmpltbank
or \trigtotmplt.
The \inspiral program then (i) conditions the data by applying an 8th order
Butterworth highpass filter with 10\% attenuation at 100~Hz and resamples the
data from 16384~Hz to 4096~Hz, (ii) acquires the current calibration
information for the data, (iii) computes an average power spectrum from the 15
analysis segments, and (iv) filters the analysis segments against the bank of
filter templates and applies the $\chi^2$ test for waveform consistency.  These
operations (apart from the data conditioning) are described in detail in the
paper discussing the \findchirp algorithm.  The \inspiral program requires
several threshold levels: the signal-to-noise threshold $\rho_\ast$, the
chi-squared threshold $\chi^2_\ast$, the number of bins $p$ to use in the
chi-squared veto, and the parameter $\delta$ used to account for the mismatch
of a signal with a template when implementing the chi-squared veto.  Several
additional parameters used by the \inspiral program include the duration
(and therefore the number) of analysis segments, the highpass filter order
and frequency, the low-frequency cutoff for the waveform templates and the
inverse spectrum truncation duration ($\pm16$~s).

The \inspiral program produces a list of inspiral triggers that have crossed
the signal-to-noise ratio threshold and have survived the chi-squared veto.
These triggers must then be confronted with triggers from other detectors to
look for coincidences.

\subsection{The \trigtotmplt program}
\label{ss:trigtotmplt}

The \trigtotmplt program \cite{lalapps} converts a list of triggers coming from \inspiral and
constructs an appropriate template bank that is optimized to minimize the
computational cost in a follow-up stage.  Only those templates that were found
in one detector need to be used as filters in the follow-up analysis, and only
the data segments of the second detector where triggers were found in the data
from the first detector needs to be analyzed.  Thus, in the S2 triggered search
pipeline, only the L1 dataset needs to be fully analyzed; only a small fraction
of the H1 and H2 datasets needs to be analyzed in the follow-up, and then only
a fraction of the template bank needs to be used.  The template bank produced
by \trigtotmplt is read by the \inspiral program that is run on the second
detector's data.

\subsection{The \inca program}
\label{ss:inca}

The \emph{in}spiral \emph{c}oincidence \emph{a}nalysis, \inca \cite{lalapps}, performs several tests for
consistency between triggers produced by \inspiral output from analyzing
data from two detectors.  Triggers are said to be \emph{coincident} if they
have consistent start times (taken to be $\pm1$~ms for coincidence between
triggers from H1 and H2 and taken to be $\pm11$~ms for coincidence between
Hanford and Livingston triggers, where 10~ms is the light travel time
between the two sites and 1~ms is taken to be the timing accuracy of our matched
filter).  The triggers must also be in the same waveform
template and the measured effective distance, $D_{\mathrm{H1}}$ and
$D_{\mathrm{H2}}$ in H1 and H2 respectively, of the putative inspiral must agree
in triggers from H1 and H2:
$|D_{\mathrm{H1}}-D_{\mathrm{H2}}|/D_{\mathrm{H1}}<\kappa+\epsilon/\rho_{\mathrm{H2}}$
where $\kappa$ and $\epsilon$ are tunable parameters and $\rho_{\mathrm{H2}}$
is the signal-to-noise ratio observed in H2.  Such an amplitude
consistency is not applied in the coincidence requirements between Hanford and
Livingston triggers because the difference in detector alignment at the sites,
while small, is significant enough that a significant fraction of possible signals
could appear with large amplitudes ratios.  In addition, in comparing Hanford
triggers, it is important to not discard H1 triggers that fail to be coincident
with H2 triggers merely because H2 was not sensitive enough to detect that
trigger.  Therefore H1/H2 coincidence is performed as follows:
(1) For each H1 trigger compute
$(1-\kappa)D_{\mathrm{H1}}/(1+\epsilon/\hat{\rho}_{\mathrm{H2}})$ (the lower
bound on the allowed range of effective distance) and
$(1+\kappa)D_{\mathrm{H1}}/(1-\epsilon/\hat{\rho}_{\mathrm{H2}})$ (the upper
bound of the range).  Here $\hat{\rho}_{\mathrm{H2}}$ is the anticipated
signal-to-noise ratio of a trigger in H2 given the observed trigger in H1.
(2) Compute the maximum
range of H2 for the trigger mass.  (3) If the lower bound on the effective
distance is
further away than can be seen in H2, keep the trigger (do not require
coincidence in H2).  (4) If the range of H2 lies between the lower and upper
bounds on the effective distance, keep the trigger in H1.
If an H2 trigger is found
within the interval, store it as well.  (5) If the upper bound on the distance
is less than the range of H2, check for coincidence.
If there is no coincident trigger discard the H1 trigger.

\section{Construction of the search pipeline}
\label{s:pipeline}

In this section we demonstrate how the S2 triggered search pipeline in
Fig.~\ref{f:pipeline} can be abstracted into a DAG to execute the analysis. We
illustrate the construction of the DAG with the short list of science segments
shown in Table~\ref{t:fakesegslist}. For simplicity, we only describe the
construction of the DAG for zero time lag results. (An artificial delay in the
trigger time can be introduced to produce false coincidences---i.e.,
coincidences that cannot arise from astrophysical signals---as a means to
obtain information on the background event rate. We refer to this as
non-zero time-lag results.) The DAG we construct filters
more than the absolute minimum amount of data needed to cover all the double
and triple coincident data, but since we were not computationally limited
during S2, we chose simplicity over the maximum amount of optimization that
we could have used.

\begin{table}[t]
\begin{center}
\begin{tabular}{cccc}
\hline
Interferometer&Start Time (s)&End Time (s)&Duration (s)\\
\hline
\hline
L1 &  730000000 &730010000 &  10000  \\
L1 &  731001000 &731006000 &   5000  \\
L1 &  732000000 &732003000 &   3000  \\
\hline
H1 &  730004000 &730013000 &   8000  \\
H1 &  731000000 &731002500 &   2500  \\
H1 &  732000000 &732003000 &   3000  \\
\hline
H2 &  730002500 &730008000 &   5500  \\
H2 &  731004500 &731007500 &   2500  \\
H2 &  732000000 &732003000 &   3000  \\
\hline
\end{tabular}
\end{center}
\caption[Fake Science Segments Used to Test DAG Generation]{
\label{t:fakesegslist}
The fake science segments used to construct the DAG shown in figure
\ref{f:fake_segs_dag}.  Start and end time is given in GPS time
(seconds since 0h UTC 6~Jan 1980).
}
\end{table}

A DAG consists of \emph{nodes} and \emph{edges}. The nodes are the programs
which are executed to perform the inspiral search pipeline. In the S2 triggered
search pipeline, the possible nodes of the DAG are \datafind, \tmpltbank,
\inspiral, \trigtotmplt, and \inca.
The edges in the DAG define the relations between programs; these relations
are determined in terms of \emph{parents} and \emph{children}, hence the
directed nature of the DAG. A node in the DAG will not be executed until all
of its parents have been successfully executed. There is no limit to the
number of parents a node can have; it may be zero or many. In order for the
DAG to be acyclic, no node can be a child of any node that depends on the
execution of that node. By definition, there must be at least one node in the
DAG with no parents. This node is executed first, followed by any other nodes
who have no parents or whose parents have previously executed.  The
construction of a DAG allows us to ensure that inspiral triggers for two
interferometers have been generated before looking for coincidence between the
triggers, for example.

The S2 inspiral DAG is generated by a script which is an implementation of the
logic of the S2 triggered search pipeline.  The script reads in all science
segments longer than $2048$~seconds and divides them into \emph{master analysis
chunks}. If there is data at the end of a science segment that is shorter than
$2048$~seconds, the chunk is overlapped with the previous one, so that the
chunk ends at the end of the science segment.  An option is given to the
\inspiral code that ensures that no triggers are generated before this time and
so no triggers are duplicated between chunks.  Although the script generates
all the master chunks for a given interferometer, not all of them will
necessarily be filtered. Only those that overlap with double or triple
coincident data are used for analysis.  The master analysis chunks are
constructed for L1, H1 and H2 separately by reading in the three science
segment files.  For example, the first L1 science segment in
Table~\ref{t:fakesegslist} is divided into the master chunks with GPS times
730000000--730002048,
730001920--730003968,
730003840--730005888,
730005760--730007808,
730007680--730009728, and
730007952--730010000 (with triggers starting at time 730009664).

The pipeline script next computes the disjoint regions of double and triple
coincident data to be searched for triggers. 64 seconds is subtracted from the
start and end of each science segment (since this data is not searched for
triggers) and the script performs the correct intersection and unions of the
science segments from each interferometer to generate the segments
containing the times of science mode data to search.
The L1/H1 double coincident segments are
730007936--730009936 and 731001064--731002436;
the L1/H2 double coincident segments are
730002564--730004064 and
731004564--731005936; and
the L1/H1/H2 triple coincident segments are
730004064--730007936 and 732000064--732002936.
The GPS start and end times are thus given for each segment to be searched for
triggers.  The script uses this list of science data to decide which master
analysis chunks need to be filtered. All L1 master chunks that overlap with H1
or H2 science data to be searched are filtered. An L1 template bank is
generated for each master chunk and the L1 data is filtered using this bank.
This produces two intermediate data products for each master chunk, which are
stored as XML data. The intermediate data products are the template bank file,
\verb|L1-TMPLTBANK-730000000-2048.xml|, and the inspiral trigger file,
\verb|L1-INSPIRAL-730000000-2048.xml|. The GPS time in the filename
corresponds to the start time of the master chunk filtered and the number
before the \verb|.xml| file extension is the length of the master chunk.

All H2 master chunks that overlap with the L1/H2 double coincident data to
filter are then analyzed. For each H2 master chunk, a triggered template bank
is generated from L1 triggers between the start and end time of the H2 master
chunk. The triggered bank file generated is called
\verb|H2-TRIGBANK_L1-730002500-2048.xml|, where the GPS time corresponds to
start time of the master H2 chunk to filter. All L1 master chunks that overlap
with the H2 master chunk are used as input to the triggered bank generation to
ensure that all necessary templates are filtered.  The H2 master chunks are
filtered using the triggered template bank for that master chunk to produce H2
triggers in files named \verb|H2-INSPIRAL_L1-730002500-2048.xml|. The GPS
start time in the file name is the start time of the H2 master chunk.

All H1 master chunks that overlap with either the L1/H1 double coincident data
or the L1/H1/H2 triple coincident data are filtered. The bank and trigger
generation is similar to the L1/H2 double coincident case. The triggered
template bank is stored in a file names
\verb|H1-TRIGBANK_L1-730004000-2048.xml| and the triggers in a file named
\verb|H1-INSPIRAL_L1-730004000-2048.xml| where the GPS time in the file name
is the GPS start time of the H1 master chunk. The H2 master chunks that
overlap with the L1/H1/H2 triple coincident data are described below.

For each L1/H1 double coincident segment to search, an \inca process is run to
perform the coincidence test. The input to \inca is all L1 and H1 master chunks
that overlap the segment to search. The GPS start and stop times passed to
\inca are the start and stop times of the double coincident segment to search.
The output is a file names \verb|H1-INCA_L1H1-730007936-2000.xml|. The GPS
start time in the file name is the start time of the double coincident
segment.  A similar procedure is followed for each L1/H2 double coincident
segment to search. The output files from \inca are named
\verb|H2-INCA_L1H2-731004564-1372.xml|, and so on.

For each L1/H1/H2 triple coincident segment, an \inca process is run to create
the L1/H1 coincident triggers for this segment. The input files are all L1 and
H1 master chunks that overlap with the segment. The start and end times to
\inca are the start and end times of the segment. This creates a file named
\verb|H1-INCA_L1H1T-730004064-3872.xml|
where the GPS start time and duration in the file name are those of the triple
coincident segment to search.  For coincidence between L1 and a Hanford
interferometer, we only check for time and mass coincidence (the amplitude cut
is disabled).

For each H2 master chunk that overlaps with triple coincident data, a triggered
template bank is generated. The input file to the triggered bank generation is
the \inca file for the segment to filter that contains the master chunk. The
start and end times of the triggered bank generation are the start and end
times of the master chunk. This creates a file called
\verb|H2-TRIGBANK_L1H1-730004420-2048.xml|.  The H2 master chunk is filtered
through the \inspiral code to produce a trigger file
\verb|H2-INSPIRAL_L1H1-730004420-2048.xml|.
 
For each triple coincident segment to filter, an \inca is run between the H1
triggers from the L1H1T \inca and the H2 triggers produced by the inspiral
code. The input files are the H1 \inca file
\verb|H1-INCA_L1H1T-730004064-3872.xml| and all H2 master chunk inspiral
trigger files
that overlap with this interval.

An H1/H2 coincidence step creates two files:
\verb|H1-INCA_L1H1H2-730004064-3872.xml| and
\verb|H2-INCA_L1H1H2-730004064-3872.xml|
where the GPS start time and duration of the files are the start and duration
of the triple coincident segment.  The L1/H1 coincidence step is then executed
again to discard any L1 triggers not coincident with the final list of
H1 triggers.  The input to the \inca are the files
\verb|L1-INCA_L1H1T-730004064-3872.xml| and
\verb|H1-INCA_L1H1H2-730004064-3872.xml|.
The output are the files
\verb|L1-INCA_L1H1H2-730004064-3872.xml| and
\verb|H1-INCA_L1H1H2-730004064-3872.xml|.
The H1 input file is overwritten as it is identical to the H1 output file.

Finally, we obtain the data products of the search which contain the candidate
trigger found by the S2 triggered search pipeline in these fake segments.
For the fake segments described here, the final output files will be:

\paragraph{Double coincident L1/H1 data}
\begin{verbatim}
L1-INCA_L1H1-730007936-2000.xml L1-INCA_L1H1-731001064-1372.xml 
H1-INCA_L1H1-730007936-2000.xml H1-INCA_L1H1-731001064-1372.xml
\end{verbatim}

\paragraph{Double coincident L1/H2 data}
\begin{verbatim}
L1-INCA_L1H2-730002564-1500.xml L1-INCA_L1H2-731004564-1372.xml 
H2-INCA_L1H2-730002564-1500.xml H2-INCA_L1H2-731004564-1372.xml
\end{verbatim}

\paragraph{Triple coincident L1/H1/H2 data}
\begin{verbatim}
L1-INCA_L1H1H2-730004064-3872.xml L1-INCA_L1H1H2-732000064-2872.xml 
H1-INCA_L1H1H2-730004064-3872.xml H1-INCA_L1H1H2-732000064-2872.xml 
H2-INCA_L1H1H2-730004064-3872.xml H2-INCA_L1H1H2-732000064-2872.xml
\end{verbatim}

\begin{sidewaysfigure}[t]
\begin{center}
\includegraphics[width=\linewidth]{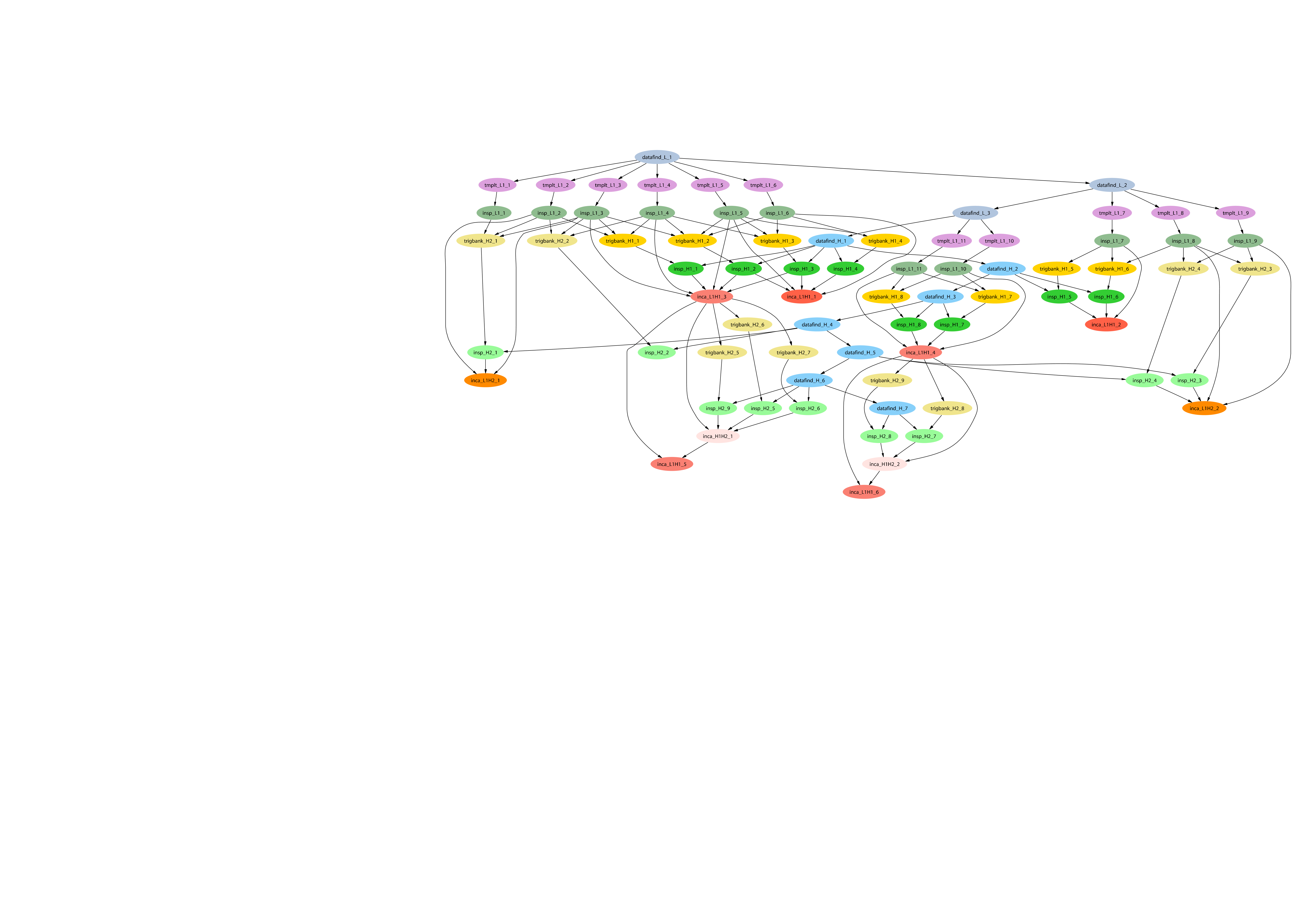}
\end{center}
\caption[DAG Generated from Fake Segments]{%
\label{f:fake_segs_dag}
The DAG generated from the pipeline shown in figure \ref{f:pipeline} and the
fake science segment list described in table \ref{t:fakesegslist}. The figure
shows the structure of the DAG with all job dependencies needed to execute the
S2 triggered search pipeline. Note that it appears that there are several Hanford
master chunks analyzed that do not need to be filtered for a zero time lag.
These are added to the DAG to ensure that all the data necessary for a
background estimation with a maximum time side of 500 seconds is analyzed.}
\end{sidewaysfigure}

The DAG that results from the fake science segments given in
Table~\ref{t:fakesegslist} is shown graphically in
Fig.~\ref{f:fake_segs_dag}.
As the size of the input science segment files increase, so the number of
nodes and vertices in the DAG increases, however the algorithm for generating
the DAG remains the same.  While the example DAG shown in 
Fig.~\ref{f:fake_segs_dag} has only 76 nodes, the analysis of the full S2 data
set required a DAG containing 6941 nodes (for the zero-lag analysis alone).
The results of the execution of this DAG to analyze the S2 data will be
reported in Refs.~\cite{s2bns,s2macho}.

\ack

The authors gratefully acknowledge the support of the United States National 
Science Foundation for the construction and operation of the LIGO Laboratory 
and the Particle Physics and Astronomy Research Council of the United Kingdom, 
the Max-Planck-Society and the State of Niedersachsen/Germany for support of 
the construction and operation of the GEO600 detector. The authors also 
gratefully acknowledge the support of the research by these agencies and by the 
Australian Research Council, the Natural Sciences and Engineering Research 
Council of Canada, the Council of Scientific and Industrial Research of India, 
the Department of Science and Technology of India, the Spanish Ministerio de 
Educacion y Ciencia, the John Simon Guggenheim Foundation, the Leverhulme Trust,
 the David and 
Lucile Packard Foundation, the Research Corporation, and the Alfred P. Sloan 
Foundation.

\end{document}